\documentclass[12 pt]{article}

\usepackage{helvet} 

\usepackage{subcaption}
\usepackage{subfig}
\usepackage{graphicx}  
\usepackage{amssymb}
\usepackage{amsmath}
\usepackage{rotating}
\newcommand{\ez}{\epsilon_0}
\usepackage{multirow}
\usepackage{soul}
\usepackage{lipsum}
              
\usepackage[margin=1.0in]{geometry}

\title{A PNP ion channel deep learning solver with local neural network and finite element input data}

\author{ Hwi Lee$^1$, Zhen Chao$^{2}$, Harris Cobb$^1$, Yingjie Liu$^1$, and Dexuan Xie$^3$ 
\thanks{Address: {\em $^1$School of Mathematics, Georgia Institute of Technology, Atlanta, GA 30332, USA.
$^2$Department of Mathematics, Western Washington University, Bellingham, WA 98225, USA.
$^3$Department of Mathematical Sciences, University of Wisconsin-Milwaukee,
Milwaukee,  WI 5320, USA.} Correspondence: {\em yingjie@math.gatech.edu, dxie@uwm.edu} } 
}

\begin{document}

\maketitle

\begin{abstract}
In this paper, a deep learning method for solving an improved one-dimensional Poisson-Nernst-Planck ion channel (PNPic) model, called the PNPic deep learning solver, is presented. In particular, it combines a novel local neural network scheme with an effective PNPic finite element solver. Since the input data of the neural network scheme only involves a small local patch of coarse grid solutions, which the finite element solver can quickly produce, the PNPic deep learning solver can be trained much faster than any corresponding conventional global neural network solvers. After properly trained, it can output a predicted PNPic solution in a much higher degree of accuracy than the low cost coarse grid solutions and can reflect different perturbation cases on the parameters, ion channel subregions, and interface and boundary values, etc. Consequently, the PNPic deep learning solver can generate a numerical solution with high accuracy for a family of PNPic models. As an initial study, two types of numerical tests were done by perturbing one and two parameters of the PNPic model, respectively, as well as the tests done by using a few perturbed interface positions of the model as training samples. These tests demonstrate that the PNPic deep learning solver can generate highly accurate PNPic numerical solutions.
\end{abstract}

\section{Introduction}
Ion channels, embedded within cell membranes, play crucial roles in many important physiological processes including nerve signaling, muscle contraction, and cell communication \cite{hille2001ion}. As one important dielectric continuum model for simulating ion transport across biological membranes, a Poisson-Nernst-Planck ion channel (PNPic) model has been developed and widely used in biophysics and computational biology to calculate macroscopic ion channel kinetics (e.g., Gibbs free energy, electric currents, transport fluxes, membrane potential, and electrochemical potential) due to its significant advantages in computational efficiency. Hence, developing its effective numerical methods can have significant impacts on ion channel studying. Mathematically, the PNPic model is a system of nonlinear partial differential equations (PDE) for computing ionic concentrations and an electrostatic potential function. Various numerical methods including finite difference \cite{bolintineanu2009poisson, zheng2011poisson}, finite element \cite{zhu2022residual, xie2023poisson, chao2021improved,xie2020finite, xie2020effective}, and integral equation methods \cite{chao2023integral} have been used to solve this model. In ion channel study and simulation, it is often required to solve the PNPic model numerically on a large mesh domain with a minuscule mesh size to explore the properties of an ion channel protein in terms of ionic concentration and electrostatic potential functions since these functions may change intensely within an ion channel pore and areas near the ion channel protein and membrane regions. However, doing so is still very expensive in numerical calculations. On the other hand, during an ion channel study, some model parameters,  interface values, and boundary values are often required to adjust properly. With a traditional numerical partial differential equation method (e.g., the finite element method), we have to resolve the model for each change of the parameters, further increasing the simulation study costs. 

Several PDE deep learning methods have been reported in the literature \cite{UniversalApprox95,liu2019multi, lkal20, nguyen2021numerical,beck2020overview}. However, they rely on a global neural network, making them have higher complexities in calculation and implementation, and demanding a large number of fine grid simulations for training data. To sharply reduce their complexities, recently, a neural network with local converging inputs (NNLCI) has been introduced in \cite{NNLCI_1d, NNLCI_2d}, opening a promising way for developing machine learning algorithms for solving one-dimensional (1D), two-dimensional (2D), and three-dimensional (3D) PDE problems. It has been shown to work for 1D and 2D Euler’s equations and electromagnetic equations, which may contain shocks, contact discontinuities, and their interactions for gas dynamics. It also has been successfully applied to predicting electromagnetic waves scattered around complicated perfect electric conductors in \cite{NNLCI_Maxwell}. In \cite{NNLCI_unstructured}, NNLCI has been developed for predicting supersonic flows in irregular domains with unstructured grids. 

NNLCI lets a trained neural network function like a local scanner which scans two coarse grid numerical solutions (with one more accurate than the other) point-wisely and then predicts the high-fidelity solution values at each scanned location. Therefore the neural network is local - taking input from local patches of the numerical solutions and generating output as the high-fidelity solution values at the center of the patches. The localness provides many nice features of the approach. Here we only list five of them:
(I) The local neural network can be small and simple. (II) The demand for training data is minimal because a fine-grid simulation can provide hundreds or thousands of local samples for training, and a couple of or tens of fine-grid simulations are usually adequate during the training process. (III) The fine-grid simulations for training can be placed sparsely (thus covering more variations of the problem) since the local samples cropped from different solutions may still bear some similarities. (IV) It can predict solutions containing discontinuities (such as shocks) sharply. (V) The training can be done on a set of computational domains and the prediction can be done on different domains.

However, it’s uncertain whether or not NNLCI will work beyond hyperbolic systems. In this work, we develop deep learning techniques to enable us to get a PNPic numerical solution in high accuracy efficiently for a range of parameters, interface values, and boundary values. This work is our first attempt to extend NNLCI to a very different PDE system with convection-diffusion equations and elliptic equations. It is also the first time for us to adapt NNLCI to deal with multiple subdomains connected with complicated interface conditions, with a distinct system of equations defined in each subdomain. With a deep learning solver, we can address two fundamental questions in a simple and elegant manor: Does NNLCI work for elliptic systems given that prior arts are for hyperbolic systems? And how to deal with multi subdomains (and different nonlinear elliptic systems) with complicated interface conditions efficiently?  

 As the first step toward the direction of developing PNPic deep learning solvers, in this work, we develop a PNPic deep learning method for solving a 1D PNPic model proposed in \cite{gardner2004electrodiffusion}. In the current literature, the majority of PNPic models for potassium channels are constructed in 1D or quasi-1D geometries. By reducing the 3D dimension to a single coordinate representing the axis along the channel pore, 1D models become computationally tractable while still maintaining basic ion channel properties. After being validated by experimental data, they are often used by biochemists, biophysicists, and bioengineers in their study of ion transport phenomena in channels. The 1D PNPic model that we select for developing PNPic deep learning methods is one important model by itself. It has been validated and applied to the simulation of ionic transport and the studies of various membrane kinetics such as membrane potentials, conductance, transport fluxes, electric current, etc \cite{roux2005ion,noskov2004control,beckstein2003liquid,peter2005ion}. It was recently selected to develop a boundary element method in \cite{chao2023integral}.  Hence, our 1D PNPic deep learning solver is valuable in biophysics, having an important impact on ion channel studies. Moreover, the simplicity of our solver bodes well for our future extension to a complicated problem involving a 3D molecular structure of an ion channel protein. 

 In particular, our PNPic deep learning solver is defined as a combination of a new 1D PNPic finite element solver with a local neural network. We adapt NNLCI here to reduce the complexity of finite element schemes when repeated simulations with varying setups of the model are conducted. We demonstrate that NNLCI works not only for hyperbolic equations as in prior works but also for the nonlinear PDE systems in multiple subdomains connected with complicated interface conditions. This is a simple feed-forward neural network that takes as input local patches of approximated solutions on two low-resolution meshes, one with a slightly higher resolution. The neural network can model the local dynamics between the solutions on the two different resolutions to predict an even more accurate solution. Thus once trained, the model can transform low-cost solutions into a high fidelity solution, by piecing together all the local predictions. The biggest cost is to generate reference solutions with the initial finite element method for training. However, compared with conventional global neural network methods \cite{UniversalApprox95,liu2019multi, lkal20, nguyen2021numerical}, our PNPic solver has a much lower dimensional input space as it only predicts on local patches. Thus, it can generate enough training data from a few training data to be accurately predictive on a broad range of very different solutions in high accuracy.
Moreover, as a new numerical method for solving a nonlinear interface boundary value problem, the novel wisdom is to assign a neural network for each subdomain and address the messy interface conditions between these subdomains. We take a bold approach --- use a single neural network to cover all subdomains, with no need to deal with any interface conditions! This simplicity of the methodology itself implies a fundamental breakthrough for solving these types of complicated problems with great robustness and efficiency for practical use. 

In this paper, we first revisit the 1D PNPic model and redefine it as a system of nonlinear variational problems. We then develop a finite element iterative algorithm for solving our 1D model and implement it as a software package in Python based on the state-of-the-art finite element library from the FEniCS project \cite{logg2012automated}, resulting in an efficient PNPic finite element solver, which is used to generate data for training and testing our local neural network as well as the PNPic reference solution required in a network training process. We next define the 1D PNPic deep learning solver as a combination of our finite element solver with a local neural network scheme such that a 1D PNPic numerical solution with a high accuracy close to the one defined in a very fine grid can be generated by using a local neural network and a set of coarse grid numerical solutions, which has a low accuracy but can be cheaply generated by a PNPic finite element solver. Remarkably, our local neural network can be trained much faster than a conventional global neural network since its input data only involves a small number of local coarse numerical solutions. Furthermore, the local neural network can work for different models disturbed by changing the parameters within a proper range provided that it is trained properly by using a database that collects numerical solutions reflecting different adjustment cases. Numerical results are finally reported, demonstrating that the PNPic deep learning solver can efficiently generate numerical solutions with a high solution accuracy.

\section{Methods}

\subsection{The Poisson-Nernst-Planck ion channel model}
\label{sec:pnp_model}

\begin{figure}[htb!]
\centering
\includegraphics[width=0.6\textwidth]{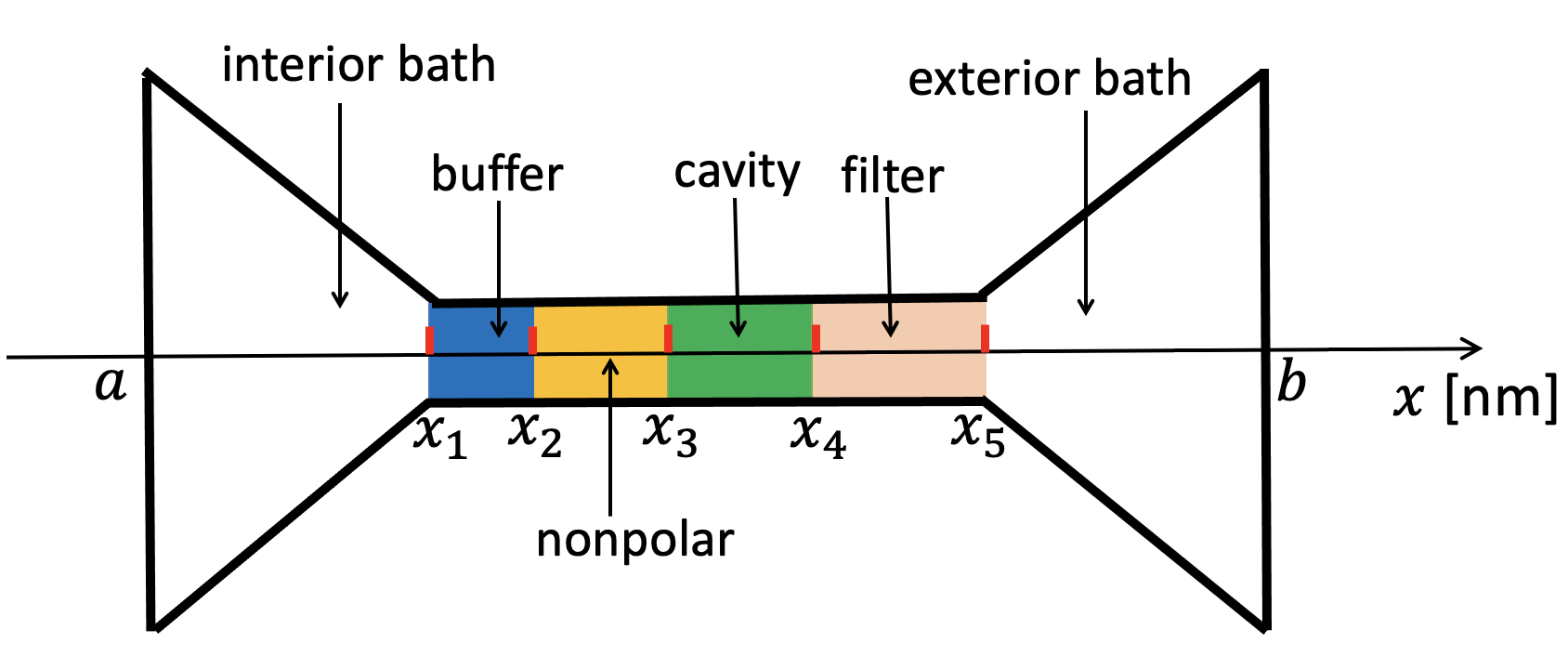}
\caption{A cross section of the K$^+$ ion channel model used in \cite{gardner2004electrodiffusion, chao2023integral}. Here the channel pore domain consists of the interior and exterior conical baths and the cylindrical channel while the channel is further split into the buffer, nonpolar, central cavity, and selectivity filter subdomains to characterize its biochemical properties.}
\label{fig:KchannelDomain}
\end{figure}

We consider a 1D PNPic model for a potassium (K$^+$) channel and a solvent with two ion species in the steady state. This model is first proposed in \cite{gardner2004electrodiffusion} and has been validated by experimental data and subsequent research that cites or builds upon that work. As needed to develop a finite element method for solving this model, we revisit it in our notation and reformulate it in a variational form. In particular, we split the channel pore domain into three regions --- an interior bath region, a channel region, and an exterior bath region. The two baths are set as cones while the channel is set as a cylinder. To reflect the biochemical features of a potassium channel in the model, we further split the channel region into four subdomains, called the buffer, nonpolar, central cavity, and selectivity filter, respectively. Because of the rotational symmetry, we can set the $x$-axis to go through the center of the channel pore such that the channel pore domain can be described as an interval, $[a, b]$, and the six subdomains can be represented by the six sub-intervals of the interval $[a, b]$ as illustrated in Figure~\ref{fig:KchannelDomain} with the five partition numbers $x_i$ for $i=1,2,3,4, 5$. The cross-sectional area $A$ of the channel pore can then be calculated by the formula $A(x)= \pi r(x)^2$ for $a\leq x \leq b$ with $r$ being a radius. Here $r$ is a constant in the channel region interval $[x_1, x_5]$, a linear function of $x$ in the interior bath region interval $[a, x_1]$, and another linear function in the exterior bath region interval $[x_5, b]$. 

Let $c_1$ and $c_2$ denote the concentration functions of species 1 and 2, respectively, and $\phi$ an electrostatic potential function of an electric field induced by charges from an ion channel protein and the solvent. They are defined by a PNPic model as follows:
\begin{subequations}
\label{PNP_K_channel}
\begin{align}
\label{PNP_K_channel_P}
-&A(x)^{-1} \frac{d}{dx}\left[\ez \epsilon(x) A(x) \frac{d\phi(x)}{dx} \right] = 
        {e_c} \left[ z_1c_1(x) + z_2 c_2(x)\right] + \rho(x), \quad a < x < b, \\[2.5pt]
\label{PNP_K_channel_NP}
& A(x)^{-1} \frac{d}{dx} \left[ A(x) D_i(x) \left( \frac{d c_i(x)}{dx} + z_{i} \frac{ e_c}{k_{B} T} c_i(x) \frac{d \phi(x)}{dx}\right)\right] = 0,  \quad a < x < b,
\end{align}
\end{subequations}
for $i=1, 2$, along with Dirichlet boundary conditions
\begin{subequations}
\label{eqn:K_channel_BC}
\begin{align}
\phi(a) &= \phi_a, \quad \phi(b) = \phi_b, \label{eqn:K_channel_P_BC}\\
c_i(a) &=c_{i,a}, \quad  c_i(b) = c_{i,b}, \quad i = 1, 2,
\end{align}
\end{subequations}
where $\ez$ is the permittivity constant of vacuum, $\epsilon$ is the permittivity function, $z_1$ and $z_2$ are the charge numbers for anions and cations, respectively, $D_i$ is a diffusion function of species $i$, $k_B$ is the Boltzmann constant, $T$ is the absolute temperature, $e_c$ is the elementary charge, $\phi_a, \phi_b$, $c_{i, a}$, and $c_{i,b}$ are given boundary values, and $\rho$ is a permanent charge density, which is used to account for the charge effect from an ion channel protein. Here $\epsilon$ and $D_i$ are differentiable and the model is measured in the International System of Units. Thus, the units of $\phi$, $c_i$, $A(x)$, and $D_i$ are volts, per cube meters, square meters, and square meters per second, respectively, while $e_c/\ez$ is about $1.8095 \times 10^{-8}$ volts meters and $k_B T/e_c$ is about 0.026 volts at $T=298.15$ Kelvin. Equation \eqref{PNP_K_channel_P} is the Poisson equation and the two equations of \eqref{PNP_K_channel_NP} are the Nernst-Planck equations. 

To properly reflect the biophysical properties of the channel pore in different subdomains, we set $\epsilon$ and $D_i$ as piecewise constant functions in the expressions:
\begin{equation}
\epsilon(x) = 
    \begin{cases}
        \epsilon_{ib}, & a\leq x < x_1, \\
        \epsilon_{b}, & x_1\leq x < x_2, \\
        \epsilon_{n}, & x_2\leq x < x_3, \\
        \epsilon_{c}, & x_3\leq x < x_4, \\
        \epsilon_{f}, & x_4\leq x < x_5, \\
        \epsilon_{xb}, & x_5\leq x \leq b,
    \end{cases}
\qquad 
\qquad
D_i(x) = 
    \begin{cases}
        D_{ib}, & a\leq x < x_1, \\
        D_{b}, & x_1\leq x < x_2, \\
        D_{n}, & x_2\leq x < x_3, \\
        D_{c}, & x_3\leq x < x_4, \\
        D_{f}, & x_4\leq x < x_5, \\
        D_{xb}, & x_5\leq x \leq b,
    \end{cases}
\end{equation}
where $ \epsilon_{ib},  \epsilon_{b},  \epsilon_{n},  \epsilon_{c},  \epsilon_{f},  \epsilon_{xb},$ $D_{ib}, D_{b}, D_{n}, D_{c},  D_{f}$, and $  D_{xb}$ are positive numbers. 
In this case, we have to formulate the PNPic model defined in ~\eqref{PNP_K_channel} as an interface boundary value problem, which involves six equations defined in the six sub-intervals and connected by interface conditions. We do not present this formulation instead of reformulating ~\eqref{PNP_K_channel} into a variational problem since we solve the PNP ion channel model by a finite element method in this work.

Multiplying each equation of \eqref{PNP_K_channel_P} and ~\eqref{PNP_K_channel_NP} by a test function, $v$, and using integration by parts, we can get the variational problem as follows:

 Find $\phi \in H^1(a, b)$ and $c_i \in H^1(a, b)$ for $i=1, 2$ satisfying the Dirichlet boundary value conditions of ~\eqref{eqn:K_channel_BC} such that
 \begin{equation}
\label{eqn:weak_p}
\begin{aligned}
    &\int_{a}^{b} \epsilon(x) A(x) \frac{d\phi(x)}{dx} \frac{dv(x)}{dx} \mathrm{~d} x -
 \frac{e_c}{\ez}  \int_{a}^{b} A(x) \left[z_1 c_1(x) + z_2 c_2(x) \right] v(x) \mathrm{~d} x \\
 =& \frac{1}{\ez} \int_{a}^{b}A(x) \rho(x) v(x) \mathrm{~d} x \quad \forall v\in H_0^1(a, b),
\end{aligned}
\end{equation}
\begin{equation}
\label{eqn:weak_np}
\int_{a}^{b} A(x) D_i(x) \left[ \frac{d c_i(x)}{dx} + z_{i} \frac{ e_c}{k_{B} T} c_i(x) \frac{d \phi(x)}{dx}\right] \frac{d v(x)}{dx} \mathrm{~d} x = 0 \quad \forall v\in H_0^1(a, b),
\end{equation}
for $i=1, 2$. Here $H^1(a,b)$ is the Sobolev function space, in which all the functions have the first weak derivatives over the interval $(a, b)$, and $H_0^1(a, b)$ is defined by
\[  H_0^1(a, b) =\{v \in H^1(a,b)  \mid v(a)=0, v(b)=0\}.\]

\subsection{The PNPic finite element iterative solver}
Let $V$ denote a linear finite element function space, which is a finite-dimensional subspace of $H^1(a,b)$ based on a mesh partition of the interval $[a, b]$ with mesh size $h$. Each function of $V$ is linear within each subinterval of the partition and continuous on the interval $[a, b]$. We define a subspace, $V_0$, of $V$ by
$$
V_0=\{v \in V \mid v(a)=0, \quad v(b)=0\}.
$$
Substituting $V$ and $V_0$ to  $H^1(a,b)$ and  $H_0^1(a,b)$, respectively, we immediately derive a system of finite element equations from the variational equations of \eqref{eqn:weak_p} and~\eqref{eqn:weak_np}. We propose to solve this system by a damped iterative scheme as defined below.

Let $\phi^{(k)}$ and $c_i^{(k)}$ denote the $k$-th iterates of $\phi$ and $c_i$, respectively. When the initial iterates $\phi^{(0)}$ and $c_i^{(0)}$ are given, we define the updates $\phi^{(k+1)}$ and $c_i^{(k+1)}$ by
\begin{equation}
\label{eqn:fem_ite}
\begin{aligned}
& c_i^{(k+1)}=c_i^{(k)}+\omega\left(p_i-c_i^{(k)}\right), \quad i=1,2, \\
& \phi^{(k+1)}=\phi^{(k)}+\omega\left(q-\phi^{(k)}\right),
\end{aligned}
\end{equation}
where $\omega$ is a damping parameter between 0 and 1, $p_i$ is a solution of the linear finite element variational problem: Find $p_i \in V$ satisfying $p_i(a)=c_{i, a}$ and $p_i(b)=c_{i, b}$ for $i=1,2$ such that for $i=1, 2,$
\begin{equation}
\label{eqn:fem_np}
\int_{a}^{b} A(x) D_i(x) \left[ \frac{d p_i(x)}{dx} + z_{i} \frac{ e_c}{k_{B} T} p_i(x) \frac{d \phi^{(k)}(x)}{dx}\right] \frac{d v(x)}{dx} \mathrm{~d} x = 0 \quad \forall v \in V_0,
\end{equation}
and $q$ is a solution of the linear finite element variational problem: Find $q \in V$ satisfying $q(a)=\phi_a$ and $q(b)=\phi_b$ such that
\begin{equation}
\label{eqn:weak_p2}
\begin{aligned}
\int_{a}^{b} \epsilon(x) A(x) \frac{d q(x)}{dx} \frac{dv(x)}{dx} \mathrm{~d} x  &=
 \frac{e_c}{\ez} \int_{a}^{b} A(x) \left[z_1 c_1^{(k+1)}(x) + z_2 c_2^{(k+1)}(x) \right] v(x) \mathrm{~d} x  \\
 & \quad +\frac{1}{\ez} \int_{a}^{b}A(x) \rho(x) v(x) \mathrm{~d} x \quad \forall v\in V_0,
\end{aligned}
\end{equation}

By default, the initial iterates $\phi^{(0)}$  and $c_i^{(0)}$ are set by
\begin{equation}
    \label{initial_iterates}
   \phi^{(0)} = \frac{\phi_b - \phi_a}{b-a} x + \phi_a, \quad c_i^{(0)}=\frac{c_{i,b} - c_{i,a}}{b-a} x + c_{i,a}, \quad  i=1,2.
\end{equation} 

We continue to carry out the iterative process \eqref{eqn:fem_ite} for $k=0, 1, 2, \ldots$ until the  iteration termination rule holds:
\begin{equation}
    \label{Ite-stop}
    \|   \phi^{(k+1)}  -  \phi^{(k)}  \| < \epsilon,  \quad  \max_{i=1,2} \|   {c}^{(k+1)}_i -  {c}^{(k)}_i  \| < \epsilon,
\end{equation} 
where $\epsilon$ is a tolerance (e.g. $\epsilon=10^{-6}$) and  $ \| \cdot \|$ is defined by  
\[     \|v\| = \sqrt{\int_a^b |v(x)|^2 dx}. \]

We implemented the above finite element iterative solver as a Python software package based on the state-of-the-art finite element library from the FEniCS project \cite{logg2012automated}. In the implementation, we scaled $\phi$ as a dimensionless potential function by the formula $\frac{e_c}{k_BT}\phi$. Following \cite{gardner2004electrodiffusion}, we used nanometers, moles per liter, and $10^{-5}$ cm$^2$ per second as the units of length, concentration, and diffusion, respectively. We also used the approximation values of some related physical constants (e.g. $e_c/(k_BT)$) given in \cite{gardner2004electrodiffusion}. In this way, we could produce the same test results as those reported in \cite{gardner2004electrodiffusion}. Furthermore, we validated our PNPic solver and package by using an analytical solution of the PNPic model \eqref{PNP_K_channel}. See Appendix for details. 

\subsection{The PNPic deep learning solver}
Following the approach of NNLCI \cite{NNLCI_1d}, we define our PNPic deep learning solver as a combination of our PNPic finite element solver with a local neural network scheme.  We start its definition with a uniform coarse grid mesh of interval $[a,b]$, 
\[ a=x^0<x^1<\dots <x^{N_1}< x^{N_1 +1}=b,\]
where the $i$th grid point $x^i=a+i h_1$ with the mesh size $h_1=(b-a)/(N_1+1)$. We refine the grid to obtain another uniform coarse grid with the mesh size $h_{2}=h_1/2$. We then solve the two nonlinear finite element systems of equations ~\eqref{eqn:weak_p} and~\eqref{eqn:weak_np} derived from these two coarse grids by our PNPic finite element software, respectively, producing two low cost PNPic finite element solutions as the input data for the local neural network scheme.

The architecture of the neural network is a simple fully connected feed-forward network with the activation function being set as $\tanh$. As done in \cite{NNLCI_1d}, the neural network scheme acts like a local post processor to scan these two coarse grid finite element solutions and refine them into a high-fidelity solution of the PNPic model pointwisely (i.e., with a much better accuracy than the input from coarse grid solutions.) In this neural network scheme, we use a small neural network consisting of an input layer with 20 neurons, three hidden layers, each of which contains 15 neurons, and the output layer with 3 neurons. 

We now construct $N_1$ input data sets, denoted by $U_i$ for $i=1,2, \ldots, N_1$, with each set having 20 elements from the two coarse grid finite element solutions. A PNPic solution contains three functions, $\phi$, $c_1$, and $c_2$. For clarity, let $u$ denote one of them. At the $i$th mesh point $x^{i}$ of the first coarse grid mesh, the first part of the input data set $U_i$ contains the three function values of $u$ at the mesh points $x^{i-1}, x^i$, and $x^{i+1}$, respectively, which are denoted as
\begin{equation}
    u_{i-1}, u_{i}, u_{i+1}, 
\end{equation}
and the second part of the input data set $U_i$ as
\begin{equation}
    u_{i'-2}, u_{i'}, u_{i'+2},
\end{equation}
which, respectively, denote the values of $u$ at the mesh points $x^{i'-2}, x^{i'}$, and $x^{i'+2}$ of the second coarse grid at the same locations of the mesh points $x^{i-1}, x^i$, and $x^{i+1}$ of the first coarse grid. To this end,  the input data set $U_i$ is obtained in the form
\begin{equation}
\label{input_system}
\begin{array}{ll}
U_i =& \{\phi_{i-1}, \phi_{i}, \phi_{i+1},c_{1,i-1}, c_{1,i}, c_{1,i+1},c_{2,i-1}, c_{2,i}, c_{2,i+1}, h_1,\\
&\;\; \phi_{i'-2}, \phi_{i'}, \phi_{i'+2},c_{1,i'-2}, c_{1,i'}, c_{1,i'+2},c_{2,i'-2}, c_{2,i'}, c_{2,i'+2}, h_2 \}.
\end{array}
\end{equation}
Note that the grid sizes $h_1$ and $h_2$ of the two coarse grids have been included in the input data set $U_i$. This treatment comes from \cite{NNLCI_unstructured} to enable the neural network solver to work for irregular grids. 

In the neural network training process, we define the loss function by
\begin{equation}
\label{loss_function}
\begin{aligned}
 {\text{Loss}}= \displaystyle \sum_{sample}\sum_{i=1}^{N_1} \|&(\mbox{output of the deep learning solver at input data $U_i$})\\
& - ({\text{reference solution at }}  x^i) \|^2,
\end{aligned}
\end{equation}
where $\|\cdot \|$ is the $2$-norm of vectors. This loss function measures the difference between the output and reference solutions at the first coarse grid points $x^i$ for $i=1,2,\ldots, N_1$ if we are interested in the prediction of a set of PNPic solution values at the first coarse grid points. Here a reference solution is a PNPic finite element solution in high accuracy, which we can generate by our solver on a very fine grid mesh. The summation of the loss function includes the input data sets corresponding to not only different spatial locations but also to different models produced from the model by using disturbed boundary conditions, disturbed parameter values or disturbed subdomains. Each of these model solutions is referred to as a sample for training in the first summation of \eqref{loss_function}. Every output of the deep learning solver and the corresponding reference solution in the loss function must be at the same spatial location for the same sample. 

In the training process, we start with multiple steps of Adams optimizer and then switch to the limited-memory BFGS (L-BFGS) optimization algorithm for minimizing the loss function to determine the weight and bias parameters of the neural network optimally. By default, we set the learning rates as $10^{-4}$ and $10^{-5}$ for Adams and L-BFGS, respectively. The neural network of our PNPic deep learning solver is implemented on Google Colab using Pytorch 2.1.0.

In summary, our PNPic deep learning solver is a novel combination of our efficient PNPic finite element solver with our effective local neural network scheme. Its input data sets come from the two PNPic coarse grid finite element solutions and it outputs a predicted PNPic solution in very high accuracy. Compared with the traditional finite element method, which can only find one approximate solution of the model, our PNPic deep learning solver can predict a family of PNPic solutions with high accuracy for a family of PNPic model problems provided that it is well trained by using the PNPic solution data generated from perturbed parameters, perturbed subdomains, and perturbed boundary values, etc.    

\subsection{Numerical test results}

To demonstrate that our PNPic deep learning solver can be used to produce an approximate solution of the PNPic model in high accuracy for a family of model problems, we did different numerical tests in this section.
Following \cite{gardner2004electrodiffusion}, we set the interval $[a, b]$ with $a=-5$ and $b=8.5$. We then set the interface numbers $x_1=0$, $x_2=0.2$, $x_3=1.3$, $x_4=2.3$, and $x_5=3.5$. For clarity, we list default parameter values and coefficient functions of the PNPic model in Table~\ref{tab:K_channel}. From this table, we can get that the radius function  $r(x)$ is a constant function with value 0.5 in $0 \leq x \leq 3.5$, a linear function in the interior bath, defined by the two values $r(-5) =5.5$ and $r(0)=0.5$, and another linear function determined by $r(3.5)=0.5$ and $r(8.5) = 5.5$ in the exterior bath. These intervals and parameter values come from a study of the K$^+$ channel \cite{gardner2004electrodiffusion,chao2023integral}.
We construct two coarse grid meshes with the mesh sizes $h_1=1/40$ and $h_{2}=1/80$ and one fine grid mesh with the mesh size $h_{ref}=1/640$. These meshes include the interface points $x_i$ for $i=1,2,3, 4, 5$ as mesh grid points. We solve the PNPic model \eqref{PNP_K_channel} by our software package based on these three meshes. 

Following \cite{gardner2004electrodiffusion}, we set the charge density function $\rho$ within each subdomain of the ion channel pore in the expression
$$ \rho = \frac{Q}{A(x)\ell}, $$ 
where $Q$, $\ell$, and $A(x)$ denote the charge amount (in Coulomb) within the subdomain, the subdomain length, and the subdomain cross-sectional area function, respectively.

To assess the numerical solution accuracy improvement done by our PNPicDLS, we define a reduction factor, $\tau_i$, for the case of perturbation $i$ by
\begin{equation}
\label{improvement}
\tau_i = \frac{ \|U^{i}_p - U^{i}_{ref}\|}{\|U^{i}_{h2} - U^{i}_{ref}\|}, \quad i \in {\cal T},   
\end{equation} 
where  ${\cal T}$ denotes a set of perturbation cases done in the prediction/validation tests, $U^{i}_{p}$ denotes a solution of the PNPic model predicted by our PNPic deep learning solver in the $i$th perturbation, $U^{i}_{h2}$ is a numerical solution of the PNPic model on the second coarse grid mesh for training in the $i$th perturbation, and $U^{i}_{ref}$ is the numerical solution on the finest grid for the $i$th perturbation. The smaller the $\tau_i$ value, the more improvement is done by our PNPic deep learning solver in comparison with the input solution data produced by our PNPic finite element solver. Our choice of the norm in the reduction factor is the $l_2$-norm which is consistent with the loss function \eqref{loss_function} used in training.

\begin{table}[htb!]
\caption{
The subdomain intervals of a K$^+$ channel pore and the model parameters used in database generation.}
\label{tab:K_channel}
\centering
\begin{tabular}{l|c|c|c|c|c|c|c}
\hline
& Interval & $\ell$  & $r(x)$ & $A(x)$   & $Q$ &   $\epsilon$ &   $D_i$ \\
\hline
Interior bath  & [-5.0, 0.0]  & 5 & Linear function & $\pi r(x)^2$         & 0          & 80             & 1.5 \\
Buffer         & [0.0, 0.2]  & 0.2 & 1/2 & $\pi$/4     & -4$e_c$      & 80            & 0.4 \\
Nonpolar       & [0.2, 1.3] & 1.1 & 1/2  & $\pi$/4     & 0               & 4             & 0.4 \\
Central cavity & [1.3, 2.3] & 1  & 1/2  & $\pi$/4     & -1/2$e_c$     & 30            & 0.4 \\
Selectivity filter & [2.3, 3.5] & 1.2 & 1/2 & $\pi$/4     & -3/2$e_c$       & 30            & 0.4 \\
Exterior bath  & [3.5, 8.5] & 5   & Linear function  & $\pi r(x)^2$    & 0               & 80             & 1.5 \\
\hline
\end{tabular}
\vskip 5pt
\end{table}

We then perform two types of perturbation tests. 

\begin{figure}[htb!]
    \centering
     \begin{subfigure}[b]{0.3\textwidth}
                \centering
                \includegraphics[width=\textwidth]{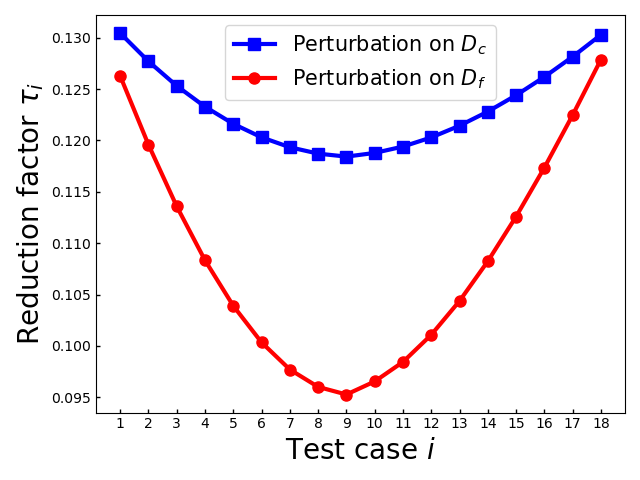}
                \caption{On diffusion $D_i$}
        \end{subfigure}
    \quad
     \begin{subfigure}[b]{0.3\textwidth}
                \centering
                \includegraphics[width=\textwidth]{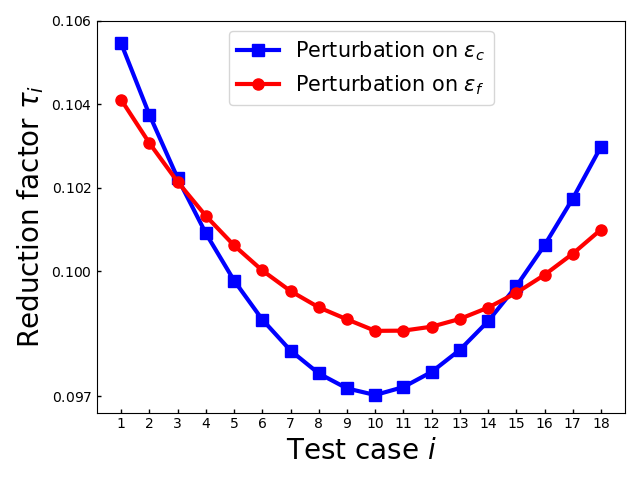}
                \caption{On permittivity $\epsilon$}
        \end{subfigure}
    \quad
     \begin{subfigure}[b]{0.3\textwidth}
                \centering
                \includegraphics[width=\textwidth]{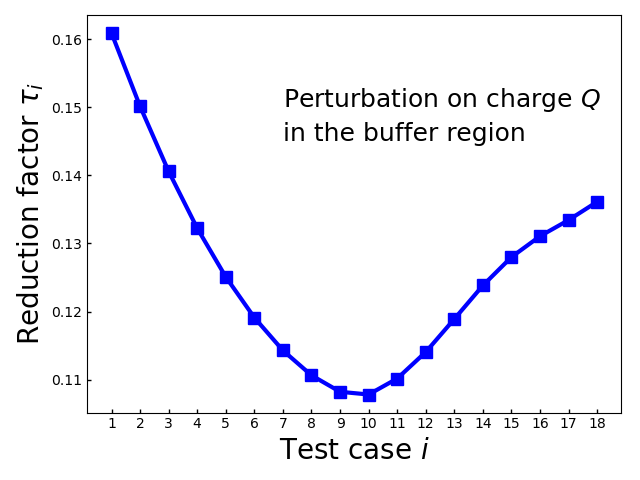}
                \caption{On charge $Q$}
        \end{subfigure}
    \caption{Numerical solution accuracy improvements by our PNPic deep learning solver in terms of the reduction factor $\tau_i$ defined in \eqref{improvement}. Here 18 prediction test cases were done by our PNPic deep learning solver for the five parameters $D_{c}, D_{f}$, $\epsilon_c$, $\epsilon_f$, and $Q$, respectively. }%
    \label{fig:charge_diffusion_permittivity}%
\end{figure}

In the first type of tests, we perturb only one parameter, say $p$, of the PNPic model from $p$ to $\hat{p}$ with $\hat{p}=(1+\lambda)p$ for $\lambda=0, \pm 0.01, \pm 0.02, \ldots, \pm 0.09, \pm 0.1$ while retaining all the other parameters. That is, a total of $21$ tests are carried out and in each test we generate three different PNPic finite element solutions based on the two coarse grid meshes and the fine grid mesh. We then split the generated data set into the training data set that includes the solution data from the tests using $\hat{p}=(1+\lambda)p$ for $i = 0, \pm 0.1$  and the testing data set that includes the solution data from the tests using $\hat{p}=(1+\lambda)p$ for $\lambda=\pm 0.01, \pm 0.02, \ldots,  \pm 0.08, \pm 0.09$. In our numerical tests, we set $p$ as a value of $D_c, D_f, \epsilon_c, \epsilon_f$, or $Q$ as follows:
\[ D_c = 0.4, \quad D_f=0.4, \quad  \epsilon_c = 30, \quad \epsilon_f = 30, \quad Q=-4e_c,\]
where $Q$ is the total charge amount in the buffer region. In these tests, we used five different neural networks for the five single parameter perturbation cases. We calculated the reduction factor $\tau_i$ of \eqref{improvement} for each test case and reported these values in Figure~\ref{fig:charge_diffusion_permittivity}. From this figure we can see that all the values of the reduction factor $\tau_i$ were less than $0.17$, indicating that our predictions have high accuracy. 

In the second type of test, we perturb two parameter values simultaneously. Let $p_1$ and $p_2$ denote those two parameters to be disturbed from their default values in Table~\ref{tab:K_channel}. We then disturb them to $(1+\delta_1)p_1$ and $(1+\delta_2)p_2$ for $ \delta_1 = 0,\pm 0.1$ and $\delta_2 = 0,\pm 0.1$. The two parameters selected for tests are the two permittivity constants in the cavity and filter regions, $\epsilon_c$ and $\epsilon_f$, the two diffusivity constants in the cavity and filter regions, $D_c$ and $D_f$, and the two boundary values of the two concentrations $c_1$ and $c_2$, $c_{i, a}$ and $c_{i,b}$ with $a=-5$ and $b=8.5$ for $i=1,2$. As done in the case of single parameter case, we dedicate one neural network to each choice of two parameter values, hence the total of 4 neural networks. The specific testing cases and reduction factors are reported in Tables \ref{tbl:two_parameters_results_1} and \ref{tbl:two_parameters_results_2}. 

\begin{table}[htb!]
\begin{center}
\begin{tabular}{||c| c |c | c || c| c| c| c||} 
 \hline
  Test case $i$ & $\epsilon_{c}$ & $\epsilon_{f}$ &  $\tau_i$  &  Test case $i$ & $D_{c}$ & $D_{f}$ &  $\tau_i$ \\  
  \hline 
  1 & $28$ & $32$ & 0.0996     & 5 & $0.36$ &$0.38$ &  0.1246    \\  \hline
  2 & $29$ &$29$ & 0.0994    & 6 & $0.39$ &$0.42$ & 0.1167    \\ \hline 
  3 & $31$ & $32$ & 0.0987   & 7 & $0.43$ &$0.37$ &   0.1106 \\ \hline
  4 & $33$ & $28$ &  0.0972 &  8 & $0.41$ & $0.41$ &    0.1299\\ \hline
\end{tabular}
\end{center}
\caption{Accuracy improvement by our PNPic solver subject to perturbations in permittivity and diffusivity in cavity and filter; all the other parameter values remain unchanged from those in Table \ref{tab:K_channel}. }
\label{tbl:two_parameters_results_1}
\end{table}

\begin{table}[htb!]
\begin{center}
\begin{tabular}{||c| c |c | c || c| c| c| c||} 
 \hline
  Test case $i$ & $c_{1,a}$ & $c_{1,b}$ &  $\tau_i$  &  Test case $i$ & $c_{2,a}$ & $c_{2,b}$ &  $\tau_i$ \\  
  \hline 
  9 & $0.142$ & $0.140$ & 0.1521     & 13 & $0.141$ &$0.148$ &  0.1143    \\  \hline
  10 & $0.137$ &$0.153$ & 0.1159    & 14 & $0.138$ &$0.157$ & 0.1364    \\ \hline 
  11 & $0.157$ & $0.144$ & 0.1196   & 15 & $0.152$ &$0.148$ &   0.1052 \\ \hline
  12 & $0.163$ & $0.162$ &  0.1634 &  16 & $0.162$ & $0.155$ &    0.1413\\ \hline
\end{tabular}
\end{center}
\caption{Accuracy improvement by our PNPic solver subject to perturbations in boundary conditions of the two concentrations; all the other parameter values remain unchanged from those in Table \ref{tab:K_channel} . }
\label{tbl:two_parameters_results_2}
\end{table}

\begin{figure}[htb!]
    \centering
    \begin{subfigure}[b]{0.48\textwidth}
                \centering
                \includegraphics[width=\textwidth]{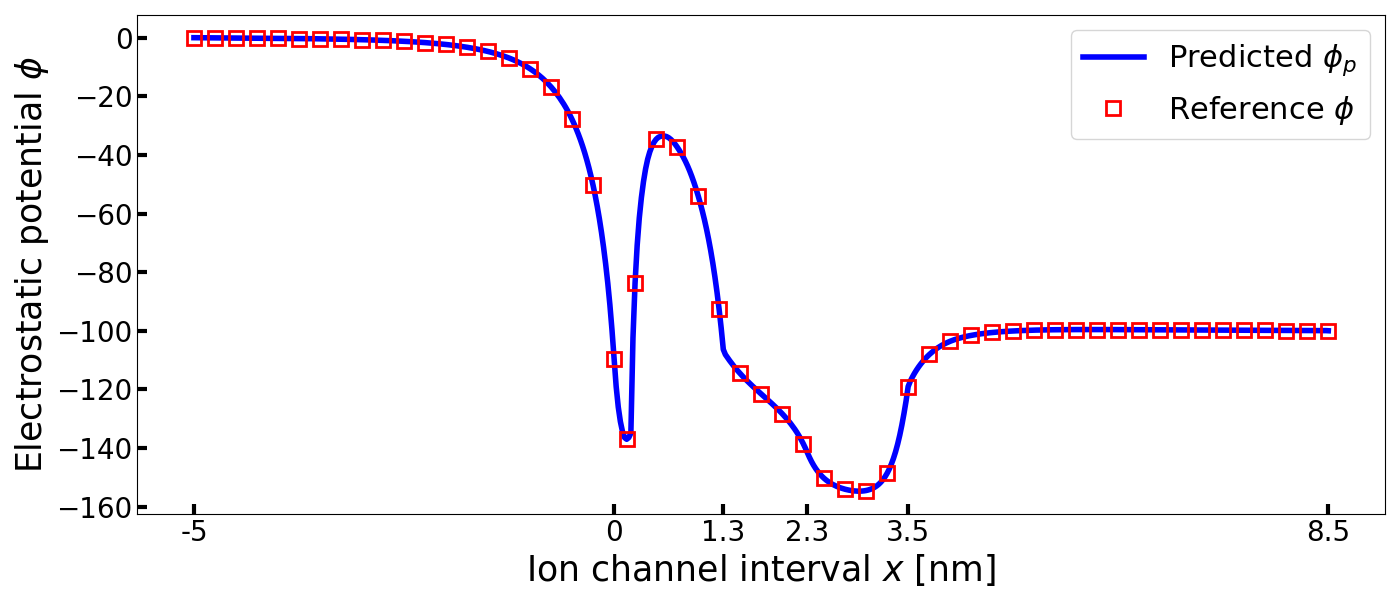}
                \caption{On charge $Q$}
        \end{subfigure}
    \quad
    \begin{subfigure}[b]{0.48\textwidth}
                \centering
                \includegraphics[width=\textwidth]{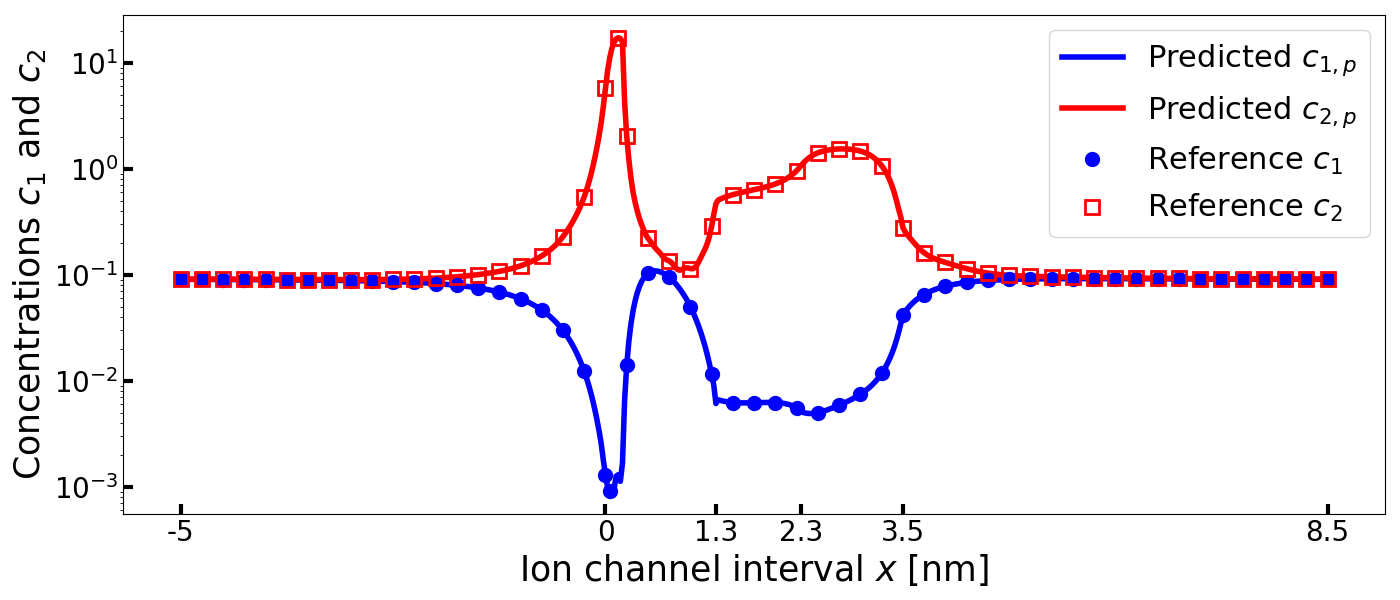}
                \caption{On charge $Q$}
        \end{subfigure}
    \caption{Comparison of the predicted solution functions $\phi_p$, $c_{1,p}$, and $c_{2,p}$ of the PNPic model by our PNPic deep learning solver with the reference solution functions $\phi$, $c_{1}$, and $c_{2}$ generated by our PNPic finite element software package on the fine grid mesh with mesh size $h_{ref}=1/640$ for the test case with $-7\%$ perturbation of the permittivity constant $\epsilon_c$ in the cavity region.}%
    \label{fig:test1_sol_plot}%
\end{figure}

\begin{figure}[htb!]
    \centering
    \begin{subfigure}[b]{0.3\textwidth}
                \centering
                \includegraphics[width=\textwidth]{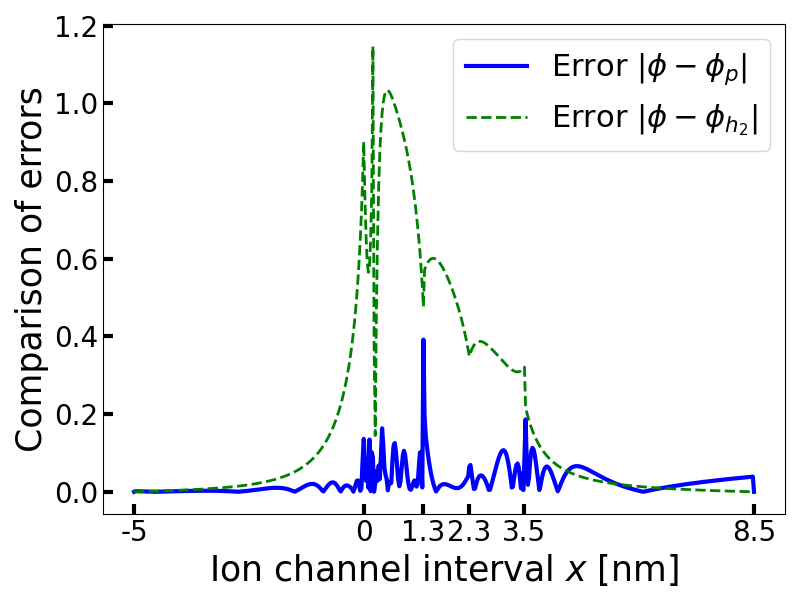}
                \caption{Potential $\phi$ case}
        \end{subfigure}
         \quad
        \begin{subfigure}[b]{0.3\textwidth}
                \centering
                \includegraphics[width=\textwidth]{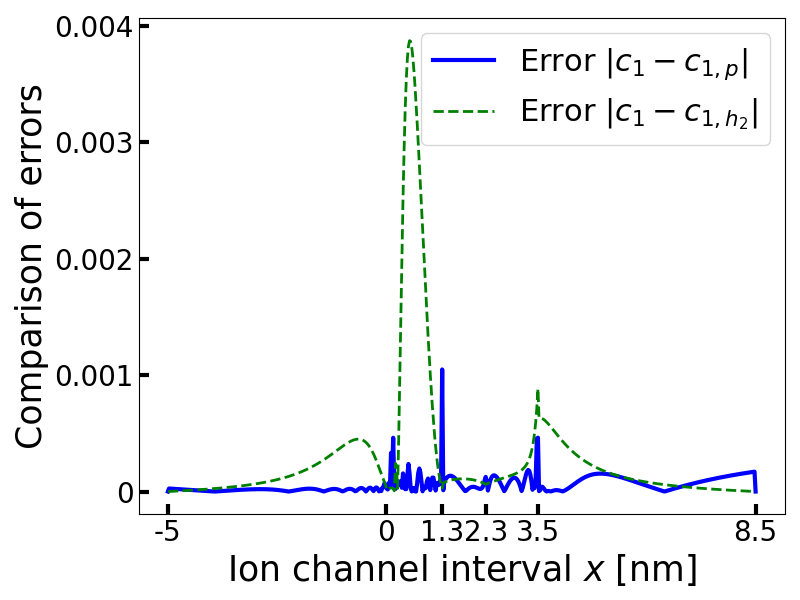}
                \caption{Concentration $c_1$ case}
        \end{subfigure}
         \quad
        \begin{subfigure}[b]{0.3\textwidth}
                \centering
                \includegraphics[width=\textwidth]{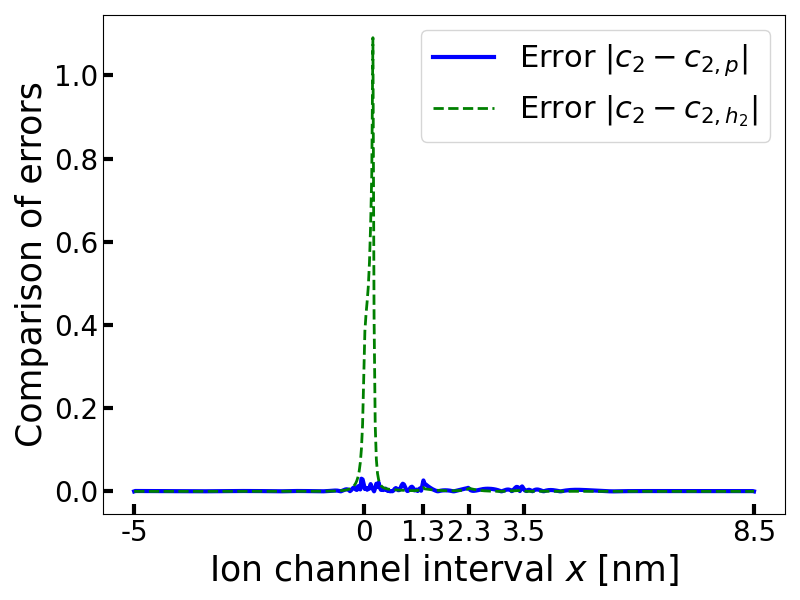}
                \caption{Concentration $c_2$ case}
        \end{subfigure}
        \caption{Comparison of the error functions $|\phi(x) - \phi_p(x)|$ and $|c_i(x) - c_{i,p}(x)|$ for $i=1,2$ of the predicted solution functions $\phi_{p}$ and $c_{i,p}$ by our PNPic deep learning solver with the error functions $|\phi(x) - \phi_{h_2}(x)|$ and $|c_i(x) - c_{i,h_2}(x)|$ of the input solution functions $\phi_{h_2}$ and $c_{i,h_2}$ generated by our PNPic finite element software package on the coarse grid mesh with mesh size $h_2=1/80$. Here $\phi$ and $c_i$ denote the reference solution functions, which we generated by our PNPic finite element software package on the fine grid mesh with mesh size $h_{ref}=1/640.$ }%
    \label{fig:test1_error_plot}%
\end{figure}

In our final numerical test, we illustrate that our PNPic deep learning solver can also accurately predict solutions corresponding to shifted interface locations. To this end, we adjusted the interface number $x_5 = 3.5$, which separates the filter and exterior bath regions, by adding the perturbations $\pm 4 h_1, \pm 3 h_1, \pm 2 h_1, \pm h_1$. Our training data set consists of the cases using $-4 h_1, -h_1, 2h_1, 4h_1$. In contrast with the previous experiments, we decreased the training gap (relative difference between adjacent two parameters in reference solutions used for training) from $10\%$ to approximately $5\%$ due to solution sensitivity on the interface location. Our particular selection of the training data set is guided by comparison of the coarse grid input solutions to take into account the various effects of perturbing the interface location on the solutions. The results for the testing cases (including the case for the unperturbed interface location) are reported in Table \ref{tbl:int_loc}.

\begin{table}[htb!]
\begin{center}
\begin{tabular}{||c| c| c| c |c |c ||} 
 \hline
  &  \multicolumn{5}{c ||}{Perturbation amount on the interface number $x_5$} \\ \cline{2-6}
  & $-3h_1$ & $-2h_1$  & $0$  & $+h_1$ & $+3h_1$   \\ 
 \hline
 $\tau_i$ &      0.1950 &   0.2041   & 0.2053   & 0.1878  &  0.2287\\ 
 \hline
\end{tabular}
\end{center}
    \caption{Accuracy improvement of our PNPic deep learning solver for testing cases with the interface located at $x_5 + p$ for $p=-3h_1, -2h_1, 0, h_1, 3h_1$. Here $\tau_i$ is the reduction factor of Case $i$ defined in \eqref{improvement}. }
    \label{tbl:int_loc}
\end{table}

\section{Discussion}

We can see from Figure \ref{fig:charge_diffusion_permittivity} that our PNPic deep learning solver can produce a more accurate numerical solution of the PNPic model than the input solution computed by the PNPic finite element solver. Given that the underlying finite element solver is second order accurate, the reduction factor of $0.17$ amounts to greater accuracy than what would be expected if we were to reduce the finer mesh size from $h_2$ to $h_2/2$. The closer the parameter values are to the default ones from Table~\ref{tab:K_channel}, the smaller the reduction factors, (larger accuracy improvement) which can potentially be attributed to the symmetry of the training data sets. 

The qualitative behaviors of the predicted solutions agree well with the reference solutions as shown in Figure \ref{fig:test1_sol_plot}. Our PNPic solver can predict the concavity and local extremes of the reference solutions. The accuracy improvement can be seen more clearly in Figure \ref{fig:test1_error_plot}. The large errors in the predicted solutions appear near the interfaces between different subdomains; our neural networks take as inputs \emph{all} local patches of solutions without differentiating whether they belong entirely to one subregion. For more complicated interfaces in multi-dimensional settings, one may consider separate, localized neural networks that are specialized for interface treatment. 

Tables \ref{tbl:two_parameters_results_1} and \ref{tbl:two_parameters_results_2} illustrate that the reduction factors for the testing cases of perturbing two parameters are comparable to the single parameter cases. The training cost becomes more expensive as the number of training data sets increases with the number of parameter values. However, the training gaps are around $5\%$ for different interface positions used in training and no smaller than $10\%$ for other parameters set for training. This allows us to use tens of fine reference solutions corresponding to different parameter values and domains to cover a relatively large parameter space ($O(1)$ in size for the convex hull of the parameters used in training.) The trained neural network can predict the high-fidelity solution for any parameter values in the parameter space.

We can see from Table \ref{tbl:int_loc} that our solver is robust with respect to perturbations in interface positions. This is particularly important for biological applications involving protein channels of different sizes and shapes. We leave it for future work to further demonstrate the performance of our PNPic solver for more complex geometry.

\section{Conclusion} 
We have developed a PNPic deep learning solver, which is defined as a combination of a new 1D PNPic finite element solver with a local neural network, and is the first to adapt NNLCI to a different PDE system with convection-diffusion equations and elliptic equations. In addition, it is the first time for us to use NNLCI to deal with multiple subdomains connected with complicated interface conditions. Instead of assigning a neural network to each subdomain and addressing the interface conditions between these neural networks, we only use one single local neural network with no need to deal with the interface conditions in the neural network. The simplicity of the methodology provides us with great robustness and efficiency for practical use, and implies a breakthrough for this type of complicated PDE problem. 

NNLCI was originally introduced for predicting weak solutions of 1D Euler systems containing shocks, contacts and their interactions in \cite{NNLCI_1d}. This 1D work paves the way for the successful application of NNLCI to the 2D Euler systems in \cite{NNLCI_2d}. The extension of NNLCI from 1D to a higher dimension is completely natural with no change in methodology. The most critical part of the work is to first demonstrate that NNLCI works for completely different types of equations in a multi-domain environment in 1D. Based on the prior works mentioned above, we have strong reasons to believe that the highly successful approach developed for the 1D PNPic deep learning solver also works for the development of a 3D PNPic deep learning method for solving a 3D PNPic model.
We will develop a 3D PNP ion channel deep learning solver for solving a 3D PNP ion channel model and train it by using the data reflecting various ion channel protein molecular structures. We expect that the 3D solver will not only capture additional spatial details and phenomena but also have orders of magnitude reduction in complexity and training costs in comparison to fine-grid simulations and conventional global neural network approaches respectively. For example, if the coarse-grid cell size is at least $8$ times as large as that of the reference-grid 
as in the paper, the number of coarse-grid cells will be 1/64 in 2D and 1/512 
in 3D of those of the reference-grid cells. This implies about two and three orders of 
magnitude reduction in complexity in 2D and 3D, respectively.  The choice between 1D and 3D PNPic deep learning methods depends on the specific research questions, available computational resources, and required simulation accuracy levels in particular study.

\section*{Acknowledgments}
This work was partially supported by the Simons Foundation through the research award number 711776 and the National Science Foundation, USA, through the award number DMS-2153376.

\section*{Author Contributions}
Lee conducted neural network training and analyzed simulation results. Chao and Lee generated data for training and prediction. Lee, Chao, Cobb, and Xie contributed to the package development. Authors 1 to 5 contributed to the manuscript writing and editing. Liu and Xie designed methodology and managed the research activities. 

%

\section*{Appendix: Validation tests by an analytical solution}

We validated our PNPic software package by using an analytical solution of the following PNP ion channel model:
\begin{subequations}
\label{PNP_model_analysis}
\begin{align*}
&-A(x)^{-1} \frac{d}{dx}\left[ \epsilon(x) A(x) \frac{d\phi(x)}{dx} \right] = 
        10 \left[ z_1c_1(x) + z_2 c_2(x)  +g(x)\right],  \quad 0 < x < 1, \nonumber \\
&A(x)^{-1} \frac{d}{dx} \left[ A(x) D_i(x) \left( \frac{d c_i(x)}{dx} + 40 z_{i} c_i(x) \frac{d \phi(x)}{dx}\right)\right] = f_i(x),  \quad i=1, 2, \quad 0 < x < 1,  \\
&\phi(0) = 0, \quad \phi(1) = -1, \quad
c_1(0) =0, \quad c_1(1) = 0, \quad c_2(0) =0,  \quad c_2(1) = 0, \nonumber
\end{align*}
\end{subequations}  
where $z_1=-1$, $z_2=1$, $A(x) = (2-x)^2$,  $D_1=1$, $D_2=1$, $\epsilon(x)=1$, and the source terms $g, f_1$, and $f_2$ are set as follows:
\begin{subequations}
    \begin{align*}
        g(x)&=\frac{3x-4}{10-5x}+x^2(x-1),\\
        f_1(x)&=\frac{2+152x-474x^2+480x^3-160x^4}{x-2}+76-474x+720x^2-320x^3,\\
        f_2(x)&=\frac{2-164x+320x^2-160x^3}{x-2}-82+320x-240x^2.
    \end{align*}
\end{subequations}
It is easy to find the analytical solution of the above system as 
\begin{equation*}
    c_1(x) = x(1-x)^2, \quad c_2(x) =x(1-x), \quad \phi(x) = x(x-2).
\end{equation*}
In this validation test, we solved this system based on five uniform meshes of the interval $[0,1]$ with the mesh size $h=1/20, 1/40, 1/80, 1/160$, and $1/320$, respectively. We then calculated an absolute error, $E(u_h, u)$, between the finite element solution $u_h$ and analytical solution $u$ for $u=c_1, c_2, $ and $\phi$ by the expression
\begin{equation} \notag
E(u_h, u)=\sqrt{\int_0^1\left|u_h(x)-u(x)\right|^2 dx}.
\end{equation}
From the finite element theory, it is known that a linear finite element method has a quadratic order of convergence rate (i.e. $O(h^p)$ with $p=2$).  The order $p$ can be approached monotonically by 
a sequence of estimations, $\{p_{u, h_k}\}$, defined by
$$
p_{u, h_k} = \frac{\ln \left(E\left(u_{h_k}, u\right) /E\left(u_{h_{k-1}}, u\right)\right)}{\ln \left(h_k / h_{k-1}\right)},
$$
where $h_k$ denotes the mesh size of the $k$th grid mesh satisfying 
\[h_1 > h_2 > \ldots > h_{k-1} > h_{k} > \ldots > 0,\] 
and $u_{h_k}$ denotes the solution defined on the $k$th mesh.
We calculated $E(u_{h_k}, u)$ and $p_{u, h_k}$ for $u=c_1, c_2, \phi$ using $h_k = 1/(2^k 10)$ for $k=1$ to 5 and listed the results in Table~A.

From Table~A, we can see that $E(u_{h_k}, u)$ approaches zero and $p_{u,h_k}$ approaches 2, monotonically, for $u=c_1, c_2,$ and $\phi$ as the grid size $h_k$ decreases from $1/20$ to $1/320$. These numerical results coincide with the finite element theory, partially validating our PNPic software package.

\vspace{0.4cm}
    \begin{tabular}{|c|c|c||c|c||c|c|}
    \hline
$h_k$& $E(c_{1,h_k}, c_1)$ & $p_{c_1, h_k}$ & $E(c_{2,h_k}, c_2)$ & $p_{c_2, h_k}$ & $E(\phi_{h_k}, \phi)$ & $p_{\phi, h_k}$ \\ \hline
1/20 &7.2927$\times 10^{-4}$ &   &2.9614$\times 10^{-4}$ & &8.9781$\times 10^{-4}$ &  \\ \hline
1/40&1.8232$\times 10^{-4}$  &1.9999&7.3417$\times 10^{-5}$ &2.0121 &2.2476$\times 10^{-4}$ &1.9980  \\ \hline
1/80&4.5532$\times 10^{-5}$ &2.0016 &1.8319$\times 10^{-5}$ &2.0028 &5.6167$\times 10^{-5}$ &2.0006  \\ \hline
1/160&1.1372$\times 10^{-5}$ &2.0014 &4.5781$\times 10^{-6}$ &2.0005 &1.4028$\times 10^{-5}$ &2.0004  \\ \hline
1/320&2.8349$\times 10^{-6}$ &2.0008 &1.1450$\times 10^{-6}$ &1.9998 &3.4948$\times 10^{-6}$ &2.0001  \\ \hline
    \end{tabular}

\vspace{0.3cm}

Table A: Validation test results produced by our PNPic software package.
    

\end{document}